\documentclass[aps,prd,twocolumn,showpacs,nofootinbib,preprintnumbers,floatfix,superscriptaddress]{revtex4-1}
\pdfoutput=1

\usepackage{amssymb}
\usepackage{graphicx}
\usepackage{amsmath}
\usepackage{url}
\usepackage{rotating}
\usepackage{dcolumn}
\usepackage{url}
\usepackage{graphicx,color}
\usepackage{hyperref}

\font\FermiSmallfont=cmssq8 scaled 1200

\def\UMDppthead#1#2#3{
\null 
\begin{center}\vskip -1.0truein{\hbox to 7.5truein {
\hfill 
\vbox to 1in {\vfill \FermiSmallfont
              \hbox{#1}
              \hbox{#2}
              \hbox{#3}
              \vfill}
}}\vskip-0.0truein\end{center}}

\begin{document}

\title{What the Milky Way's Dwarfs tell us about the Galactic Center extended excess}

\author{Ryan E.\ Keeley} \email{rkeeley@uci.edu}
\affiliation{Department of Physics and Astronomy, University of California, Irvine, Irvine, CA 92697}

\author{Kevork N.\ Abazajian} \email{kevork@uci.edu}
\affiliation{Department of Physics and Astronomy, University of California, Irvine, Irvine, CA 92697}

\author{Anna Kwa} \email{akwa@uci.edu}
\affiliation{Department of Physics and Astronomy, University of California, Irvine, Irvine, CA 92697}

\author{Nicholas L.\ Rodd} \email{nrodd@mit.edu}
\affiliation{Center for Theoretical Physics, Massachusetts Institute of Technology, Cambridge, MA 02139}

\author{Benjamin R.\ Safdi} \email{bsafdi@umich.edu}
\affiliation{Michigan Center for Theoretical Physics, Department of Physics, University of Michigan, Ann Arbor, MI 48109}

\preprint{UCI-TR 2017-11}
\preprint{MCTP 17-19, MIT-CTP 4943}

\pacs{95.35.+d,95.55.Ka,95.85.Pw,97.60.Gb}

\begin{abstract}
The Milky Way's Galactic Center harbors a gamma-ray excess that is a
candidate signal of annihilating dark matter. Dwarf galaxies remain
predominantly dark in their expected commensurate emission. In this
work we quantify the degree of consistency between these two
observations through a joint likelihood analysis.  In doing so we
incorporate Milky Way dark matter halo profile uncertainties, as well
as an accounting of diffuse gamma-ray emission uncertainties in dark
matter annihilation models for the Galactic Center Extended gamma-ray
excess (GCE) detected by the {\em Fermi Gamma-Ray Space
  Telescope}. The preferred range of annihilation rates and masses
expands when including these unknowns. Even so, using two recent
determinations of the Milky Way halo's local density leave the GCE
preferred region of single-channel dark matter annihilation models to
be in strong tension with annihilation searches in combined dwarf
galaxy analyses. A third, higher Milky Way density determination,
alleviates this tension. Our joint likelihood analysis allows us to
quantify this inconsistency.  We provide a set of tools for testing
dark matter annihilation models' consistency within this combined
dataset. As an example, we test a representative inverse Compton
sourced self-interacting dark matter model, which is consistent with
both the GCE and dwarfs.
\end{abstract}

\maketitle

\providecommand{\eqn}[1]{eqn.~(\ref{eqn:#1})}
\providecommand{\tab}[1]{Table~\ref{tab:#1}}
\providecommand{\fig}[1]{Figure~\ref{fig:#1}}
\providecommand{\zstar}{\ensuremath{z^\ast}}
\providecommand{\zmax}{\ensuremath{z_{\text{max}}}}
\providecommand{\sH}{\ensuremath{{\cal H}}}

\begin{section}{Introduction}

The Large Area Telescope aboard the {\em Fermi Gamma-Ray Space
  Telescope}, {\em Fermi}-LAT, has observed a bright excess of gamma
rays towards the Galactic Center whose presence is robust to
systematic uncertainties in the standard background templates
\cite{Goodenough:2009gk,Vitale:2009hr,Hooper:2010mq,Hooper:2011ti,
Abazajian:2012pn,Slatyer:2013pdu,Gordon:2013vta,Macias:2013vya,
Abazajian:2014fta,Daylan:2014rsa,Calore:2014xka,Abazajian:2014hsa,
TheFermi-LAT:2015kwa,Karwin:2016tsw}.
This excess has generated a great deal of interest since dark matter
(DM) annihilation models can explain three compelling coincidences in
the signal.  First, the excess' spatial morphology matches the
predictions of a generalized Navarro-Frenk-White (NFW) profile, which
is a generic prediction of cold DM
models~\cite{Navarro:1995iw,Navarro:1996gj}.  Second, the total flux
of the signal is well fit by the annihilation cross-section required
by a thermal production scenario to generate the observed cosmological
relic abundance. Third, the spectrum roughly matches the expectations
of a tens of GeV weakly interacting massive particle (WIMP)
annihilating to standard model particles.  Should the GCE turn out to
be explained by such an annihilating WIMP DM particle, it could be the
first non-gravitational evidence of DM and the first strong clue of
the particle nature of DM.

The prompt annihilation of WIMPs is not the only class of DM models
that can explain the GCE, however.  For example, a class of
self-interacting DM (SIDM) models can explain the GCE via
up-scattering of starlight that would not be seen in dwarf galaxies
\cite{Kaplinghat:2015gha, Cui:2016mhm, Lacroix:2015wfx,
  Lacroix:2014eea, Galon:2016bka}.  Specifically, this class of SIDM
particles could annihilate into electrons (as well as the other
standard model leptons) and these electrons could up-scatter the
Galactic Center's interstellar radiation field (ISRF) via the inverse
Compton (IC) process.

There are also reasonable astrophysical interpretations of the GCE.
Most notably is that the GCE can arise from a population of unresolved
millisecond pulsars (MSP)
\cite{Abazajian:2010zy,Abazajian:2012pn,Ploeg:2017vai,Fermi-LAT:2017yoi,Macias:2013vya,Mirabal:2013rba,Petrovic:2014xra,Yuan:2014rca,Yuan:2014yda,Brandt:2015ula,OLeary:2015qpx,OLeary:2016cwz}.
Specifically, observations of MSPs in globular clusters show they have
a spectrum consistent with the spectrum of the GCE.  Further, low mass
X-ray binaries (likely progenitors of MSPs) in M31 have been observed
to follow a power law radial spatial distribution, similar to the
expectations of an NFW halo \cite{Abazajian:2012pn,Yuan:2014rca}.  Other astrophysical
explanations might include more dynamic events such as cosmic-ray
injection into the Galactic Center (GC) \cite{Carlson:2014cwa,
  Gaggero:2015nsa,Cholis:2015dea,Carlson:2016iis}.  Furthermore, the
presence of the {\em Fermi} Bubbles tell us that such dynamic events
have occurred in the past, so whatever mechanism produced the {\em
  Fermi} Bubbles, could also have produced the GCE
\cite{Su:2010qj,Fermi-LAT:2014sfa,Petrovic:2014uda}.

There have arisen a number of independent avenues that each has the
potential to challenge a DM interpretation of the GCE.  One such
avenue is to look for a gamma-ray excess from other DM halos.  Such
halos include those of galaxy clusters, the limits from which have
recently been extended to be in slight tension with the GCE
\cite{Lisanti:2017qlb,Lisanti:2017qoz}, and the Milky Way's satellite
dwarf galaxies, the most recent {\em Fermi} limits appearing in
Ref. \cite{Fermi-LAT:2016uux}. Unfortunately, {\em Fermi}-LAT
observations of both of these sources, and particularly the dwarfs,
have not seen a significant complementary gamma-ray
excess.\footnote{Note, however, there has been a low-significance
  detection of a gamma-ray excess from Reticulum II and Triangulum II,
  see e.g. \cite{Geringer-Sameth:2015lua,Hooper:2015ula}, although see
  also \cite{Drlica-Wagner:2015xua}.}  In particular, this difference
between the GCE and the dwarfs has the potential to rule out certain
classes of DM interpretations of the GCE. Specifically, any minimal
model based around a two-body annihilation process (any process where
the flux is proportional to the square of the DM density) would
exhibit this same tension.

Other avenues to test whether the GCE is better explained by
annihilating DM or astrophysics is to more precisely check whether the
morphology of the excess truly follows a smooth NFW profile.  Tension
in the morphology of the GCE has arisen from the detection of an
`X-shape' residual in the {\em Fermi} data which correlates infrared
emission as seen by the WISE telescope \cite{Macias:2016nev}.  Should
this `X-shape' template account for the entirety of the GCE, it would
challenge any DM interpretation since DM annihilation would not
produce such a shape.  Measurements of the GCE being consistent with
wavelets \cite{Bartels:2015aea} or non-Poissonian fluctuations have
also been reported
\cite{Lee:2014mza,Lee:2015fea,Linden:2016rcf,Mishra-Sharma:2016gis},
which would indicate an MSP rather than DM origin, though systematic
uncertainties in such analyses remain \cite{Horiuchi:2016zwu}.
Specifically, small scale gas clouds are left out of the model used by
\texttt{GALPROP}, software used to generate the gamma-ray templates
associated with cosmic rays propagating through the Milky Way, which
could confuse any detection of point sources near the GC
\cite{Horiuchi:2016zwu}.  Though each of these lines of evidence
against a DM interpretation of the GCE have their own systematic
uncertainties, many of these systematics are independent of each
other. Arguably, these different lines of evidence add up to strongly
indicate that the GCE is astrophysical in origin.

Our focus in the present paper is to consider one aspect of this
general line of reasoning: the consistency between the GCE and the
dwarfs, and to do so with a more detailed treatment of the systematics
coming from both sides. The discussion is structured as follows.  In
section \ref{data} we discuss the background models we investigate to
understand some of the dominant sources of systematic uncertainties in
the problem. In section \ref{dm}, we discuss DM annihilation models of
the GCE.  In section \ref{models}, we discuss alternative models to
promptly annihilating DM, including astrophysical interpretations and
SIDM models.  We conclude in section \ref{conclusion}.

\end{section}

\begin{section}{Background Models and Data}
\label{data}
\begin{figure*}[t!]
\begin{center}
    \includegraphics[width=3.25 truein]{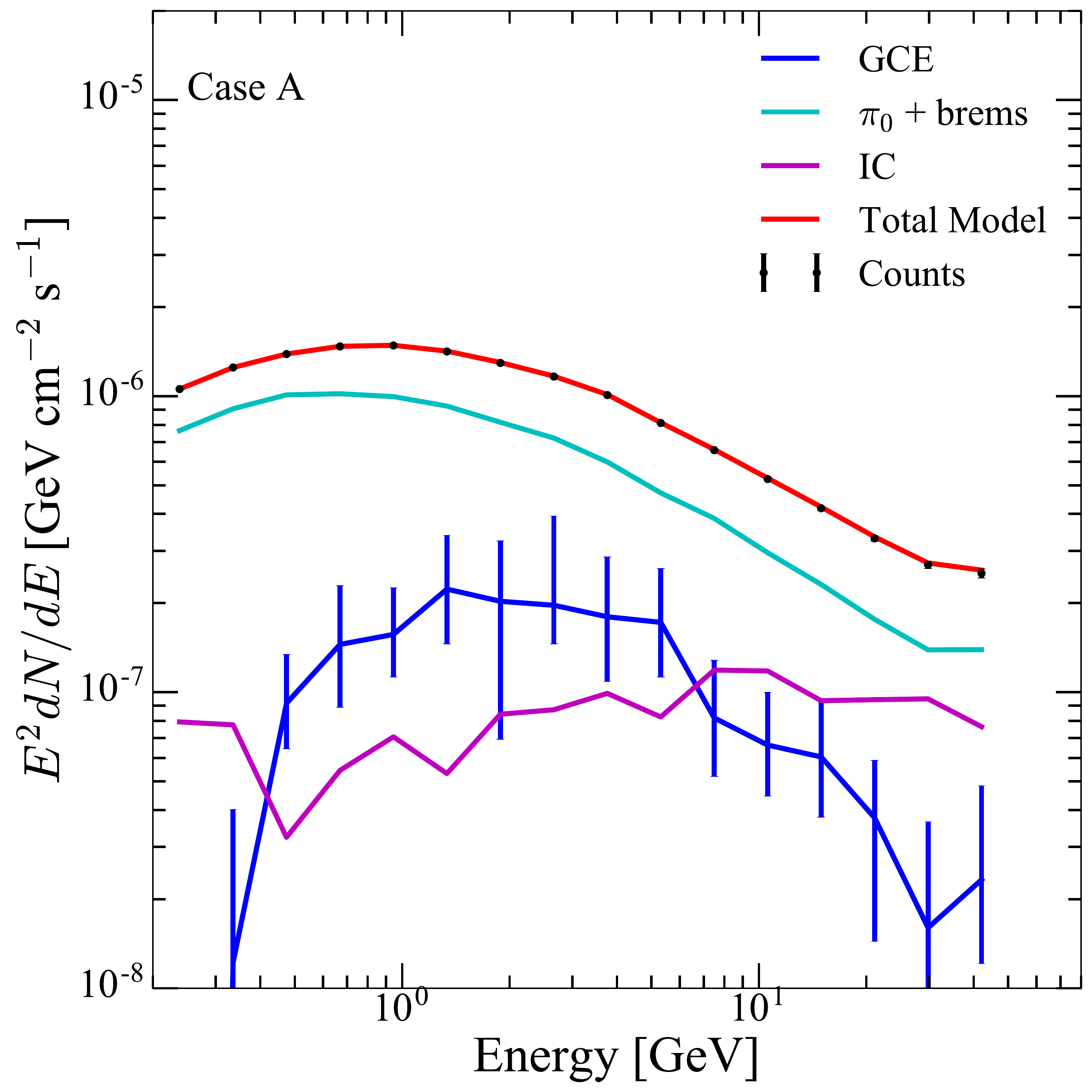}\
    \includegraphics[width=3.25 truein]{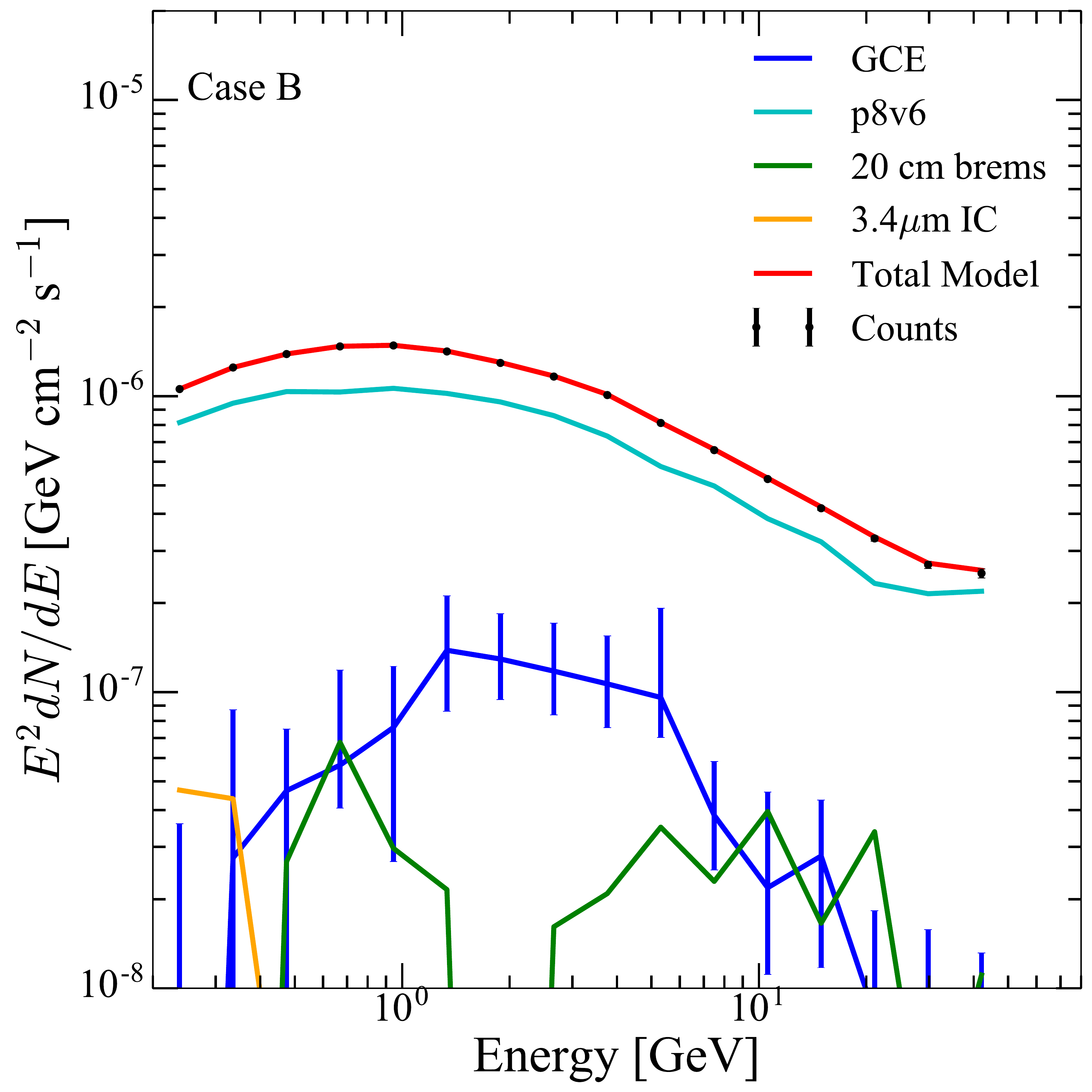}\\
    \includegraphics[width=3.25 truein]{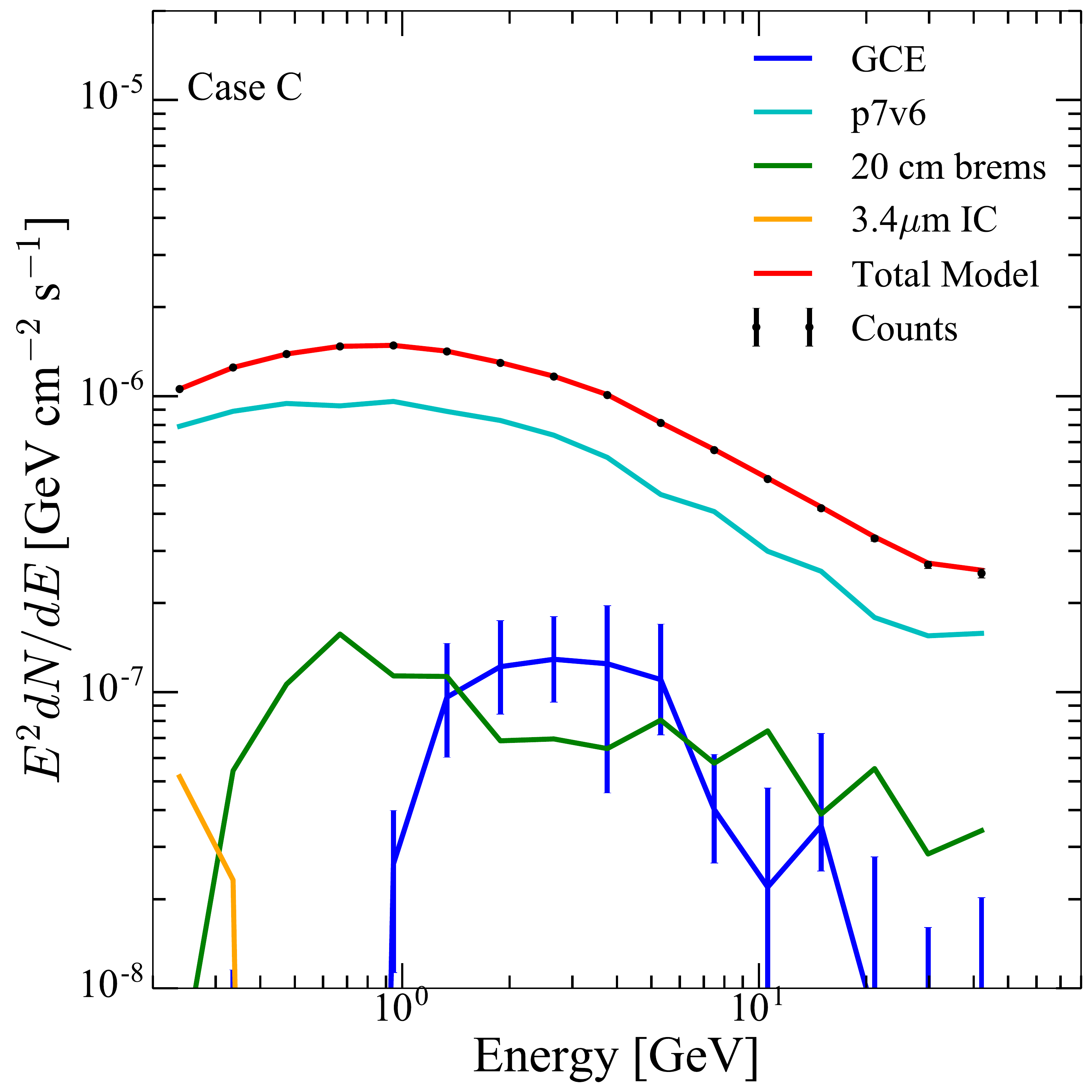}
    \includegraphics[width=3.25 truein]{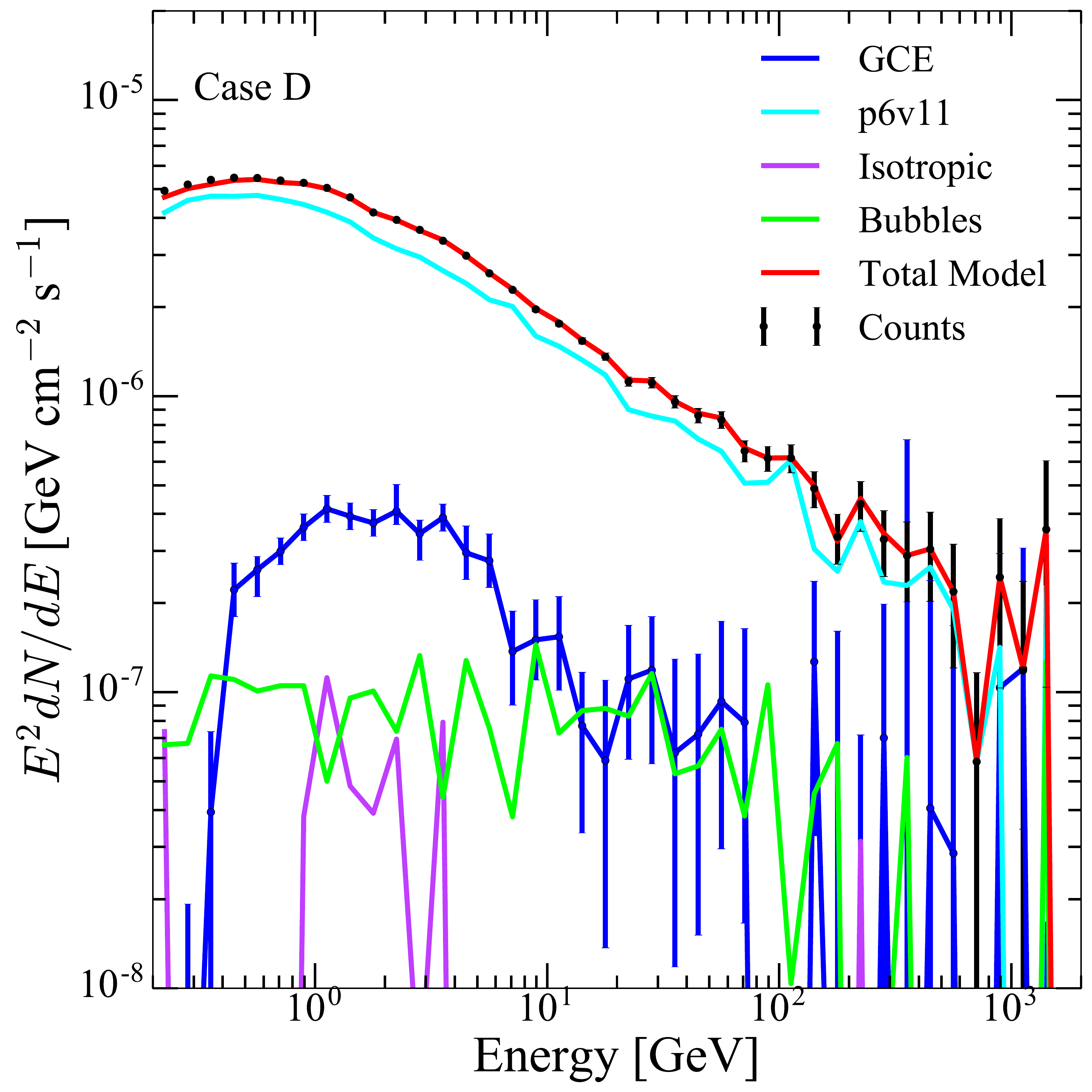}
\end{center}

\caption{Here we plot the energy flux spectrum (intensity) $E^2 dN/dE$ for the
various templates included in the likelihood fits for our A, B, C,
and D background cases.  These show the total emission from the ROI,
$7^\circ \times 7^\circ$ for cases A-C and $30^\circ \times
30^\circ$ for case D. The error bars on the counts is the Poisson error.
The various 3FGL sources were also varied in the fits but are not
included for the sake of simplicity.
\label{spectra}}
\end{figure*}

There remains significant uncertainty regarding the various processes
that contribute to the gamma-ray signal coming from the GC.  Any
interpretation of the GCE will necessarily be affected by these
uncertainties.  To capture these effects, 
we investigate four different cases of the astrophysical
background contributions to the GCE.

For all of our cases (denoted cases A, B, C, and D), we use data 
collected by {\em Fermi}-LAT.  For cases A, B, and C, that data
corresponds to observations over a 103 month period from August 2008
to March 2017.  We use all \verb|SOURCE|-class photons from the Pass 8 instrument
response functions. We apply a maximum zenith angle cut of $90^\circ$ to
avoid contamination. In cases A, B, and C, we focus our analysis on the
innermost $7^\circ \times 7^\circ$ region of interest (ROI) about the
Galactic Center.  We then bin these photons into spatial bins of size
$0.05^\circ \times 0.05^\circ$ for each energy bin. The photon events
range from 200 MeV to 50 GeV and are binned in 16 logarithmically
spaced energy bins.

For Case D we instead choose a dataset similar to that considered in 
the Inner Galaxy analyses of \cite{Daylan:2014rsa,Linden:2016rcf}. Here we 
use the best quartile, as graded by the {\em Fermi} point spread
function, of \verb|ULTRACLEANVETO|-class Pass 8 photons, gathered between 
August 4, 2008 and June 3, 2016 with recommended quality cuts. 
This case can be contrasted with the 
above in that it generally corresponds to less data, but with less 
cosmic-ray contamination and improved angular reconstruction per event. 
To mimic an earlier Inner Galaxy analyses, we use a larger ROI of
$30^\circ \times 30^\circ$, masking latitudes less than $1^\circ$.
No masks were applied to the data in cases A-C.  We also
mask the top 300 brightest and most variables sources in the 3FGL catalog
\cite{Acero:2015hja} at 95\% containment. The photons are binned into 40
equally spaced logarithmic bins between 200 MeV and 2 TeV, and spatially
using an \texttt{nside}=128 \texttt{HEALPix} grid \cite{Gorski:2004by}.

With this processed data, we perform a maximum likelihood analysis to
determine the best fit background model and GCE model. For each
component of our model, we generate a template which encodes the
spatial distribution of the photons for that component. The quantity
that we are trying to determine is then the linear combination of
these spatial templates that best fit the observed number of counts.
The templates fall into three groups: point sources, extended
emission, and diffuse emission.  The point sources we use are taken
from {\em Fermi}'s 3FGL point source catalog \cite{Acero:2015hja} and
they are typically well characterized or independent of the GCE
result. The extended emission components include a GCE template, as
well as background components coming from cosmic rays interacting with
gas in the interstellar medium (ISM) or photons in the ISRF.
Specifically, these would include any IC radiation from high energy
electron cosmic rays up-scattering the ISRF, neutral pion ($\pi_0$)
decay generated from hadronic cosmic rays interacting with the ISM,
and bremsstrahlung radiation arising from high energy electrons
interacting with the ISM. The spatial distribution of these components
are more {\em a priori} uncertain than point sources and are partially
degenerate with the GCE, especially in the lowest energy bins, where
the point-spread function is the largest.  Therefore, it is these
uncertainties and degeneracies that make a careful and broad
investigation of the diffuse backgrounds crucial to analyzing the GCE
and are the main difference between our different cases.

Since the uncertainty in the spectral shape of the GCE signal is
dominated by systematic uncertainties in the background templates,
rather than Poisson fluctuations of the total counts, it is necessary
to explore multiple possible background models.  To this end, we use
four different sets of templates for these extended background models:
\begin{itemize}
\item {\bf Case A:} We use the templates for the $\pi_0$,
  bremsstrahlung, and IC emission for `model F' from Horiuchi et
  al.\ (2016) \cite{Horiuchi:2016zwu}, which in turn used diffusion
  model parameters from Calore et al.\ (2014)~\cite{Calore:2014xka} to
  generate their background models. Their `model F' corresponds to the
  diffuse background model that was found to best fit the data in
  their ROI.  Unlike the {\em Fermi} collaboration Pass 8 and Pass 7
  diffuse backgrounds, the IC component of the diffuse background is
  fit independently of the $\pi_0$+bremsstrahlung components. We used
  the templates for `model F' from these papers. In this work, we
  denote this `case A.'
 \item {\bf Case B}: For this case, we use the Pass 8 Galactic
  interstellar emission model from the {\em Fermi} tools, which models
  the distribution of gamma rays coming from $\pi_0$ decay,
  bremsstrahlung, and IC. All three components are combined in a
  single diffuse template with fixed relative normalizations in each
  energy bin.  Furthermore, we used a template which traces the 20 cm
  radio emission first discovered by Yusef-Zadeh et
  al.\ (2013)~\cite{YusefZadeh:2012nh}.  We also include a template
  for an additional IC component that was derived from 3.4$\mu$m maps
  from the WISE telescope, discovered by Abazajian et al.\
  (2014)~\cite{Abazajian:2014hsa}.
 \item {\bf Case C:} This case uses the same templates for
  the bremsstrahlung and IC components but uses \texttt{p7v6} model
  for Pass 7.
 \item {\bf Case D:} This case uses the \texttt{p6v11} template and
   floats an isotropic template as well as a template for the
   \textit{Fermi} bubbles.
\end{itemize}

For all these cases, we allow the flux associated with each template
in a given bin to be independent of the flux in other bins, rather
than assume a specific component has a specific spectral shape. This
allows us to be agnostic about the shape of the spectrum for each of
these sources, but potentially comes at the cost of over-fitting the
data.  The results of these maximum likelihood fits for cases A-D are
shown in Fig.~\ref{spectra}.

To calculate posteriors for the dwarfs, we use the flux likelihood
limits from Albert et al.\ (2016)~\cite{Fermi-LAT:2016uux}.
Specifically, we use the flux likelihood manifolds for the nineteen
kinetically confirmed dwarf galaxies that have measured J-factors.
The J-factor for Reticulum II is calculated in Simon et al.\ (2015)
\cite{Simon:2015fdw} and the rest are calculated in Geringer-Sameth e
t al.\ (2014)~\cite{Geringer-Sameth:2014yza}.  To
calculate these flux likelihood limits, Albert et al.\ use six
years of LAT data with 24 equally-spaced logarithmic bins between 500
MeV and 500 GeV.  They binned the photons in a $10^\circ \times
10^\circ$ region about the target dwarf galaxies with a pixel size of $0.1^\circ$
in order to model any overlap from the points spread function of the
point sources in the 3FGL catalog, from the Galactic diffuse emission, 
and from the isotropic model.  Each target dwarf galaxy was modeled as 
a point-like source and used a maximum likelihood analysis with these 
templates to generate the flux likelihood limits.

\end{section}

\begin{section}{Annihilating Dark Matter Models}
\label{dm}
\begin{subsection}{Flux Spectra}
The differential flux in some ROI for the class of two-body DM
annihilation is given by the following:
\begin{equation}
\frac{d\Phi}{dE} = \frac{1}{4\pi} \frac{J}{m_\chi^2} \frac{\langle
  \sigma v \rangle}{2} \frac{dN}{dE}\,.
\end{equation}
Here, $J$ is the J-factor, the integral of the density-squared over
the ROI and through the line of sight. $m_\chi$ is the mass of the DM
particle and $\langle \sigma v \rangle$ is the thermally averaged
cross section of the annihilation. $\frac{dN}{dE}$ is the
per-annihilation spectrum, which we calculated using
\texttt{PPPC4DMID} \cite{Cirelli:2010xx}.  For our dark matter models,
we use flat priors on the DM mass and scale invariant priors on the
cross section.  The prior on the J-factor is discussed in the next
section.

\end{subsection}

\begin{subsection}{J-factors}
The J-factor is the square of the DM density integrated through the
line of sight and integrated over the ROI.
\begin{equation}
J = \int_{\rm ROI} d \Omega \int dz \ \rho^2(r(z,\Omega))\,.
\end{equation}

As in Abazajian \& Keeley 2015 \cite{Abazajian:2015raa}, we determine
the prior on the J-factor for the GC by parameterizing the
Milky Way's DM halo as a generalized NFW profile with a local
DM density ($\rho_\odot$), a scale radius ($R_s$), and an
inner profile slope ($\gamma$)
\begin{equation}
\rho(r)= \frac{\rho_\odot}{ \left( \frac{r}{R_\odot} \right)^\gamma \left( \frac{1+r/R_s}{1+R_\odot/R_s} \right)^{3-\gamma}}\,.
\end{equation}
Each of these parameters has a probability distribution, so in
principle, we could say the prior on the J-factor is the product of
the probability distributions of each of these parameters and then
perform the change of variables to write this probability distribution
as a function of the J-factor.  This is analytically cumbersome, so
we use numerical Monte Carlo methods to calculate this distribution.
Specifically, we draw values for the local density, scale radius, and
inner slope to compute a set of J-factors and then use kernel density
estimation to define the prior for the GCE J-factor.

For the local density, we use the value determined by Zhang et al.\
(2012)~\cite{Zhang:2012rsb}: $\rho_\odot = 0.28 \pm 0.08$ GeV
cm$^{-3}$. This robust determination of the local DM density
is derived from modeling the spatial and velocity distributions for a
sample of 9000 K-Dwarf stars from the Sloan Digital Sky Survey
(SDSS). The velocity distribution of these stars directly measures the
local gravitational potential and, when combined with stellar density
constraints, provides a measure of the local DM density.

The prior on the scale radius is calculated from the concentration,
which is the ratio of the virial radius to the scale radius.  The
uncertainty in the concentration is calculated from simulations of
galaxy formation.  Sanchez-Conde and Prada
(2013)~\cite{Sanchez-Conde:2013yxa} parameterized the uncertainty in the
concentration of a DM halo as a function of that halo's
mass. Thus we can write the prior on the scale radius as:
\begin{equation}
\mathcal{\log L} = -\frac{(\log_{10}(R_{\rm vir}/R_s) - \log_{10}
  c(M_{\rm vir}))^2 }{2 \times 0.14^2}\,.
\end{equation}

The prior on the inner slope we use for the Monte Carlo calculation of the
J-factor is taken to be the posterior determined by the spatial information
contained in the GCE data.  We constrain the inner slope by running the likelihood
analysis with the same background models but with different NFW
spatial templates that have different values for the inner slope.  The
likelihood analysis calculates the $\Delta$Log-likelihood value for
each of these different cases, which allows us to fit a $\chi^2$
distribution to these $\Delta$Log-likelihood values. This determines
the best fit value of $\gamma$ and its error.  Unsurprisingly, this
derived prior on the inner slope depends on the background model used.
For case A we calculate, $\gamma = 1.14 \pm 0.04$; for case B, $\gamma
= 1.24 \pm 0.04$; for case C, $\gamma = 1.10 \pm 0.04$; and for case
D, we calculate $\gamma = 1.2 \pm 0.06$.  The results of this Monte
Carlo calculation of the priors on the J-factor for the different
background cases is shown in Fig.~\ref{jfactor}.

We employ the priors on the J-factors for the dwarf galaxies from
Albert et al.\ (2016)~\cite{Fermi-LAT:2016uux}.  These are all
reported as log-normal distributions. These J-factors come with some
caveats, however.  Specifically, assumptions about how spherically
symmetric the dwarf galaxy is, which in turn can influence the
inferred cuspiness of the density profile, can lead to systematic
uncertainties greater than the statistical uncertainties
\cite{Klop:2016lug,Hayashi:2016kcy,Bonnivard:2014kza}.

\begin{figure}
\includegraphics[width=3.25 truein]{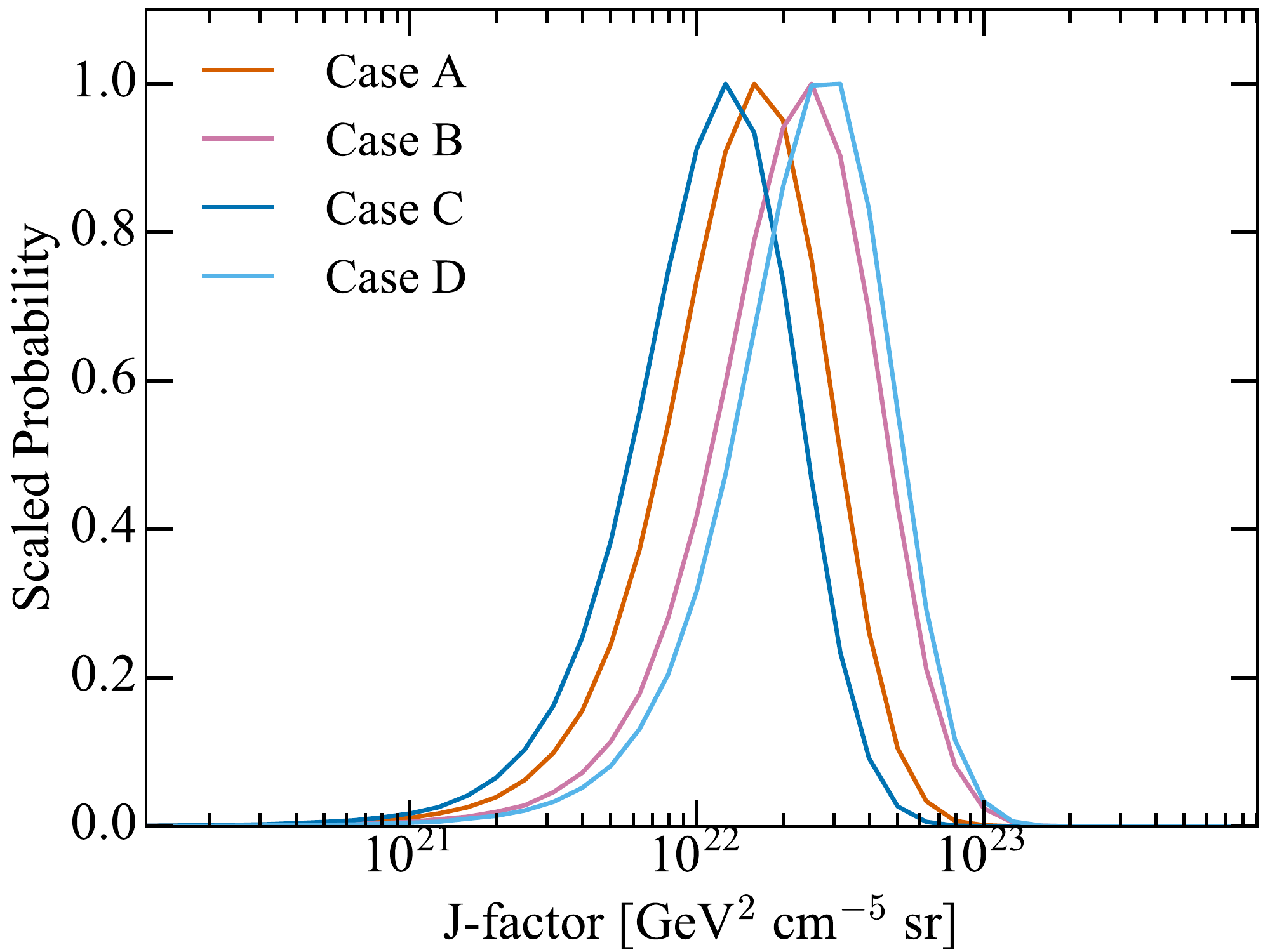}
\caption{The prior on the J-factor integrated over the ROI derived through a Monte Carlo
  convolution of the priors on the local density, scale radius, and
  inner slope.  Since each of the different background cases have different
  best fit values for $\gamma$, and since case D corresponds to a
  larger ROI, the derived uncertainties on the J-factors are
  different.
  \label{jfactor}}
\end{figure}

\end{subsection}

\begin{subsection}{Evidence Ratios}
To quantify the tension between the GCE and the dwarfs, we calculate a
Bayesian evidence ratio. This evidence ratio can be used in answering
the question: by what factor do the odds of some model being true
change with the inclusion of a new data set.  It is the product of the
Bayesian evidences of two data sets, $D_1$ and $D_2$, when considered
separately divided by the evidence of the two data sets when
considered jointly \cite{Ivesic2014}:
\begin{multline}
{\rm ER} = \frac{p(D_1) p(D_2)}{p(D_1,D_2)}\\ = \frac{\int d \theta_1
  p(D_1|\theta_1) p(\theta_1) \int d \theta_2 p(D_2|\theta_2)
  p(\theta_2)}{\int p(D_1,D_2 | \theta) p(\theta)}\,.
\end{multline}
This can be interpreted as a Bayes factor where the two models being
compared are the same except for the fact that the model corresponding
to the numerator has an additional, independent copy of the parameter
space and the two parameter spaces describe the two data sets
separately.

This statistic can indicate three different outcomes for the model.
First, if the data set $D_2$ contains no information, then this
evidence ratio is unity.  If $D_2$ is entirely consistent with $D_1$
then the evidence ratio should be less than unity. This is expected
since increasing the complexity of a model should come at a cost of
subjective belief.  If $D_2$ is in tension with $D_1$ then this
evidence ratio will be greater than unity.  How strongly this evidence
ratio prefers consistency or tension can be interpreted by any
standard Bayes factor scale.  In this work, we choose to interpret our
evidence ratios by the Jeffreys' scale.

Using this setup, we then calculate evidence ratios between the combined
dwarf galaxies and the GC. The results are stated in Table~\ref{table:ER}.

One particularly useful feature of evidence ratios in this context is
that, compared to Bayes factors, they are relatively insensitive to
systematic uncertainties in the background models.  These systematic
uncertainties can alter the total flux of the signal, but they more
drastically change in which energy bin this flux is distributed. This
is seen most clearly in the lowest energy bins, where the inclusion of
diffuse templates from 20 cm maps of bremsstrahlung emission and 3.4 
$\mu$m maps of IC emission, 
for cases B and C, remove all the photons from the NFW template for 
these bins.  Such changes to the lowest energy bins changes the overall 
curvature of the GCE spectrum, which, in turn, significantly changes the 
best fit mass but not the best fit cross section \cite{Abazajian:2015raa}.
When the best fit mass of the GCE changes, the amount of overlap between
the GCE posterior with the combined dwarfs posterior (and hence the evidence
ratio) changes relatively little.  This lack of change in overlap comes from the
fact that the contours of the dwarf posterior are almost parallel to contours of
constant cross section, since the lack of a dwarf signal contains no significant
amount of information about the spectrum.
It is because the evidence ratio is most sensitive to the cross section and not
particularly sensitive to the dark matter mass that the evidence ratios are
more robust to systematic uncertainties in the background templates.  This is
born out in Table \ref{table:ER} where the DM evidence ratios for the different
cases vary by only two orders of magnitude.  On the other hand, because the
Bayes factors are sensitive to both the normalization and the shape of the
spectrum it can vary by 30 orders of magnitude, as seen in Table \ref{reltable}.

Beyond systematic uncertainties due to the inclusion of additional
templates for bremsstrahlung and inverse Compton processes,
uncertainties in the diffuse model for the \texttt{GALPROP} generated
$\pi_0$, IC, and bremsstrahlung templates, can alter the total flux of
the GCE signal and will affect the best fit cross section for the GCE
and hence affect the tension with the dwarfs.  This is seen by the
fact the evidence ratio for our different cases significantly change.
This change is caused by the fact that the differences in these
diffuse emission templates, for cases B and D, shift the overall flux
of the GCE signal to smaller values, relative to case A. In these
background model cases, the presence of the GCE is less significant
and reduces the significance of the tension with the dwarfs.

The information encoded by the evidence ratio can be qualitatively seen
in Fig. \ref{GCE_contours}, which plots the posterior of our $b\bar{b}$
and $\tau^+ \tau^-$ DM annihilation models for each of our different GCE
background cases and for the dwarf data.  The amount of overlap in the
GCE posteriors and the dwarf posterior indicates the amount of tension
between the data sets.

For our DM annihilation models, we calculate evidence ratios between
15 and 2200 for the $b\bar{b}$ channel and between 27 and 4300 for the
$\tau^+ \tau^-$ channel.  Using the Jeffreys Scale, this indicates a
strong ($>$10) to decisive ($>$100) tension in two-body DM
interpretations of the GCE and dwarf data.

Importantly, this strong to decisive tension exists in models beyond
just the specific DM particle annihilating to $b\bar{b}$ or $\tau^+
\tau^-$.  Any model of prompt two-body decay, described by a J-factor,
would exhibit this same tension.  Hence, models that contain only
novel versions of spectrum $dN/dE$, or branching ratios, will not
alleviate this strong tension.

\begin{figure*}
\begin{center}
        \includegraphics[width=3.25 truein]{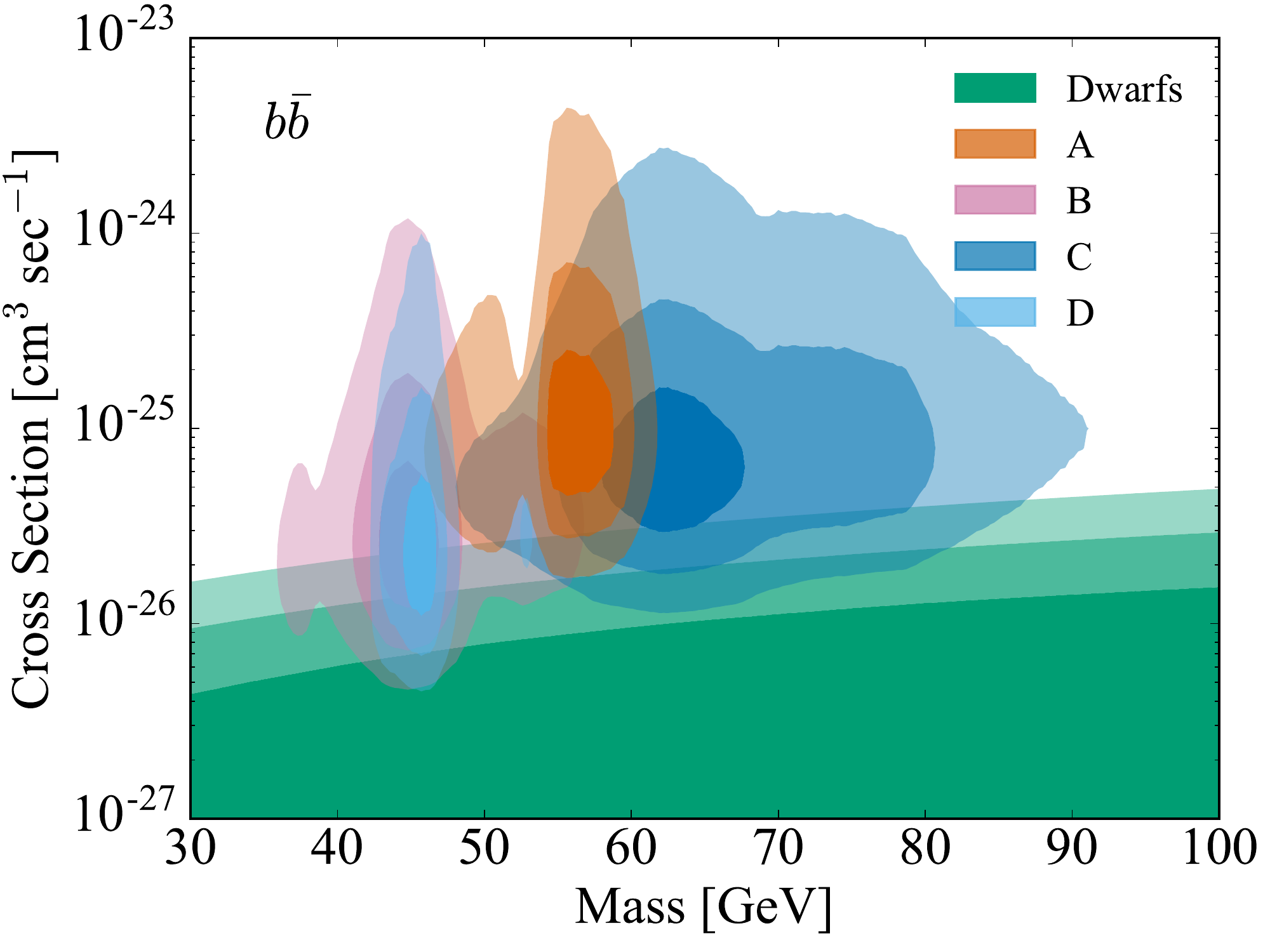}\
        \includegraphics[width=3.25 truein]{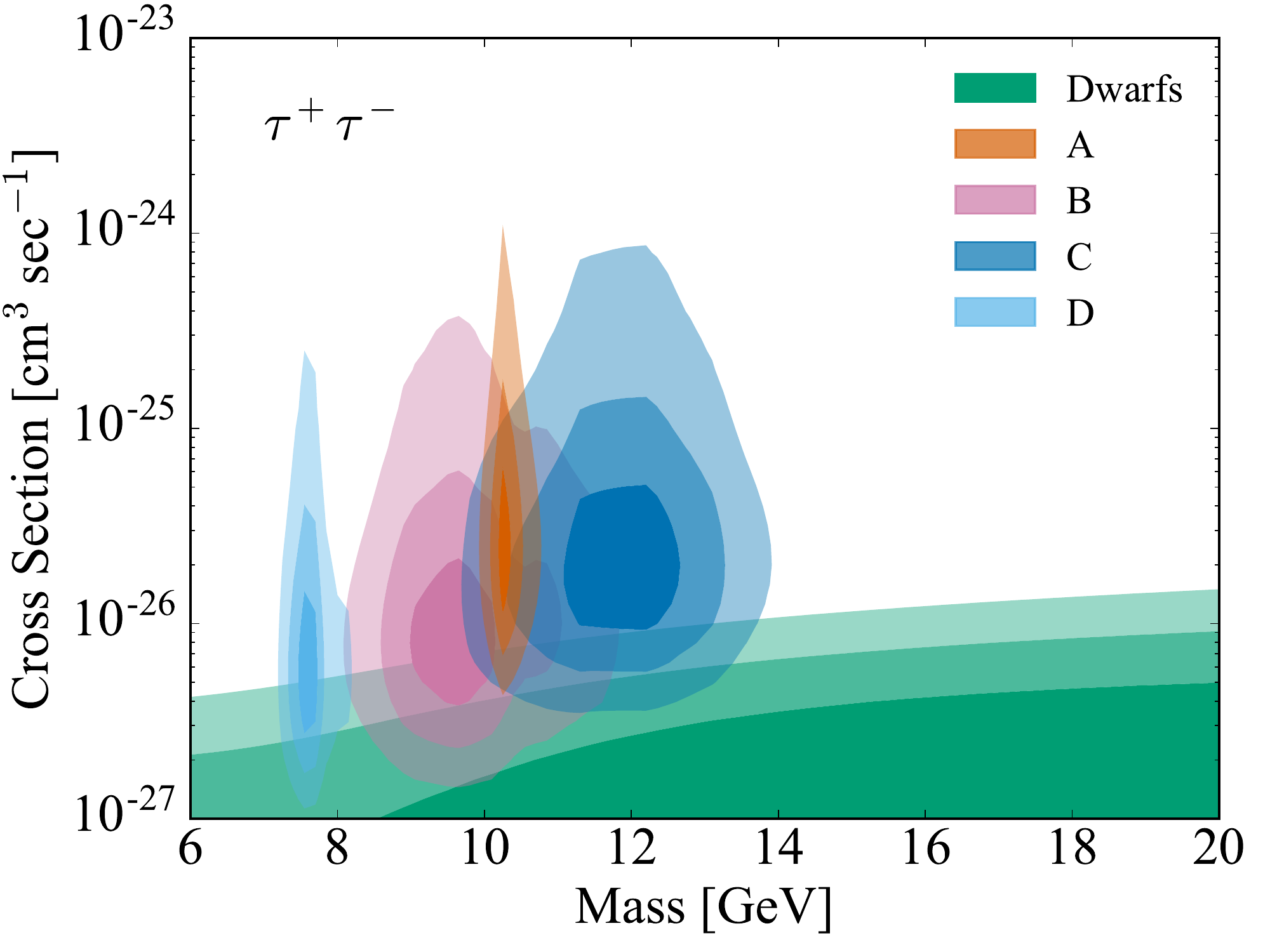}
\end{center}
    \caption{Here we show the 1, 2, and 3$\sigma$ contours of the
      posteriors for the annihilation cross section and DM
      mass.  Our calculated limits on the dwarf signal is in green,
      case A is in orange, case B is in blue, and case C is in pink.
      The results for $b\bar{b}$ on the left and $\tau^+ \tau^-$ on
      the right. The amount of overlap qualitatively demonstrates the
      information contained in the evidence ratio and shows how
      consistent two-body DM annihilation models are at explaining
      both the GCE and the lack of a dwarf
      signal.
      \label{GCE_contours}}
   
\end{figure*}

\begin{table}[]
\centering
\def\arraystretch{1.5}
\caption{Evidence ratios for our five models using the diffuse
  templates for our various background cases.}\label{table:ER}
\begin{tabular}{| l | l | l | l | l |}
 \hline
 Model                 & Case A & Case B & Case C & Case D \\ \hline
 DM: $b \bar{b}$       & 3600   & 21     & 220    & 15\\ \hline
 DM: $\tau^+ \tau^-$   & 2300   & 25     & 230    & 29\\ \hline
 Log-Parabola          & 0.69   & 0.58   & 0.71   & 0.54\\ \hline
 Exponential Cutoff    & 0.73   & 0.59   & 0.78   & 0.54\\ \hline
 SIDM                  & 1.1    & 1.2    & 1.2    & 1.1 \\ \hline
\end{tabular}
\end{table}

\end{subsection}

\begin{subsection}{Caveats}
The GCE-dwarf tension we quantified in the previous section certainly
depends on the prior information adopted for the J-factors of the GC
region and the dwarf galaxies.  Naturally, if there was a significant
change in the inferred DM content of either the GC region or
dwarf galaxies, then the nature of tension would correspondingly
change.  However, our choices for the J-factor of the GC region and
dwarf galaxies are those determined by the most robust analyses
available.

The parameter that the J-factor is most sensitive to
is the local density of DM.  As stated in a previous section,
we use a value of 0.28$\pm$0.08 GeV/cm$^3$ taken from Zhang et
al.\ (2012)~\cite{Zhang:2012rsb}.  Other groups including Pato et
al.\ (2015)~\cite{Pato:2015dua} and McKee et
al.\ (2015)~\cite{McKee:2015hwa} tend to find higher values for the
local density.  To fully resolve the tension between the GCE and the
dwarfs, the GCE J-factor needs to increase between 1 and 1.5 orders of
magnitude, which translates into a local density of 3 to 6 times
greater. As we show, none of these determinations of the local density
relieve the GCE-dwarf evidence ratio to be unity.

Another parameter with a systematic uncertainty is the scale radius of
the DM profile.  Small deviations around our fiducial value
would not change the J-factor by a great deal since the inner profile
is unchanged due to the scale radius being beyond the local radius.
However, should the scale radius become smaller than the local radius,
the inner density profile would increase as $r^{-3}$ between the local
radius and the scale radius, resulting in a larger J-factor.  A
profile with such a small scale radius could only occur in halos with
a concentration parameter far outside of what CDM simulations predict
for halos with the mass of the Milky Way.

The inner slope $\gamma$ is more robust to systematic uncertainties,
in that it is determined directly from the spatial information of the
gamma-ray data.  In particular, to fully resolve the GCE-dwarf
tension, a value of around $\gamma = 1.7$ would be required.  However,
all of the diffuse models that we tested preferred values for the
inner slope were found to be significantly below that, between and
$\gamma = 1.1$ and $1.3$.  Despite systematic uncertainties in the
parameterization of the Milky Way's DM profile, no single alteration
can fully relieve the tension between the GCE and dwarf data.

\end{subsection}

\end{section}

\begin{section}{Models}
\label{models}
We have shown that there is tension with the standard WIMP scenario
between the derived cross sections from the GC and the dwarfs, with
some important caveats.  This tension can potentially point to
alternate models being better explanations for the GCE, including
astrophysical interpretations to more complicated DM models.  To
quantitatively answer this question, we calculate a Bayes factor:
\begin{equation}
K_{12} = \frac{p(M_1|D) p(M_2)}{ p(M_2|D) P(M_1)} = \frac{p(D|M_1)}{p(D|M_2)}\,.
\end{equation}
We consider the following models: two astrophysical interpretations,
one with a log-parabola spectrum and another with an exponential cutoff
spectrum, and a SIDM model where the GCE
gamma rays are generated by DM decaying to high-energy
electrons up-scattering starlight. The Bayes factors for our models are
given in Table II.

\begin{subsection}{Astrophysical Interpretations}
Should the GCE have an astrophysical interpretation, the gamma-ray
spectrum can be parameterized as a log-parabola or a power law with an
exponential cutoff.  We investigate both parameterization as
explanations of the GCE.

The spectrum for our log-parabola model is given by:
\begin{equation}
\frac{dN}{dE} = N_0 \left( \frac{E}{E_s} \right)^{-\alpha - \beta \log(E/E_s)}\,,
\end{equation}
where $N_0$ is an arbitrary normalization, $E_s$ is a scale energy,
$\alpha$ is the slope of the power-law part of the spectrum, and
$\beta$ parameterizes the turnover of the spectrum.

The spectrum for our power law with an exponential cutoff model is
given by:
\begin{equation}
\frac{dN}{dE} = N_0 \left( \frac{E}{E_s} \right)^\gamma e^{-E/E_c}\,,
\end{equation}
where $N_0$ is the normalization of the spectrum, $E_s$ is a scale
energy, $\gamma$ is the slope of the power-law part of the spectrum,
and $E_c$ parameterizes how fast the spectrum cuts off.

Our astrophysical models do not have a specific physical
interpretation so it is not straightforward to investigate to what
extent the GCE and the lack of signal from the dwarf galaxies are
compatible given these models.  Presumably, if the GCE and any
potential dwarf signal were to be explained by the same category of
astrophysical object, then they should have the same spectral
parameters.  Therefore, it makes sense for our model to have only one
set of spectral parameters that describes both the GCE and the dwarfs.
The normalizations of the spectrum, however, would not necessarily be
the same.  One option is to allow the normalization of the spectrum of
the GCE and the spectrum of each of the dwarfs to be independent.
Following this parameterization, we calculate an evidence ratio
between the GCE and the dwarfs of about 1, which would indicate the
two data sets contain no new information relative to each other.  This
is expected, since if we put in the fact that the signals are
independent, we should get out that they have no mutual information.
Instead of saying these normalizations are entirely independent of
each other, we use a zeroth order ansatz to parameterize the
normalization as the product of the stellar mass of the system and the
gamma-ray rate per stellar mass.  The stellar mass would, of course,
be independent between regions, but the gamma-ray rate per stellar
mass should be the same between regions.  To this end, we find $N_0$
in the above equations such that the integral of $dN / dE$ over our
energy range (200 MeV to 50 GeV) is one.  This allows us to attach
physical interpretations to our normalization for $d \Phi /dE$.

Specifically, it makes sense, should the initial mass function of some
galaxy be independent of the stellar mass of that galaxy, that the
gamma-ray production rate scales linearly with the stellar mass of the
galaxy.  Hence, the gamma-ray rate per stellar mass should be
consistent across all regions.

Ultimately, this leads to the following parameterization of the
differential number flux:
\begin{equation}
\frac{d \Phi}{dE} = \frac{\dot{N}}{4 \pi R^2} \frac{M_*}{M_0} \frac{dN}{dE}\,,
\end{equation}
where $M_*$ is the stellar mass of the object, $R$ is the distance to
the object, and $\dot{N}/M_0$ is the gamma-ray rate per stellar mass,
which should be the same between different objects.

For both spectra of astrophysical models, we marginalized over the
spectral parameters with flat priors, and marginalized over the over
the gamma-ray rate per stellar mass with a scale invariant prior.  We
use the stellar mass of the dwarfs, the distance to them, as well as
the uncertainties in those parameters from McConnachie (2012)
\cite{McConnachie:2012aj}.  Interestingly, both of our astrophysical
models pick out values for the gamma-ray rate per stellar mass around
$10^{31\pm 1}$ s$^{-1}$ $M_\odot^{-1}$, which is consistent with
known millisecond pulsars.  In the end, the evidence ratios for each
of our two spectral choices for astrophysical models, for all of our
background cases, are less than unity.  Importantly, this less than
unity evidence ratio indicates that the combined dwarf and GCE data
have a weak indication of a mutual astrophysical excess described by a
single set of parameters.

The Bayes factors we compute also point towards a preference for these
astrophysical models. As seen in Table \ref{reltable}, the
log-parabola spectrum is preferred over any DM model in each of the
cases, and the exponential cutoff spectrum is preferred in three out
of four of the cases.  The preference in the Bayes factor can be
thought of as coming from two distinct sources.  One is the GCE data
on their own prefer that model and the other is that the model can
better explain the differences in the flux from the GC and the dwarfs.
Astrophysical interpretations, with evidence ratios less than unity,
can do better on the latter count, but interestingly, depending on the
data case, can also do better on the former count.  In all cases, the
log-parabola spectrum can explain the GCE data better than dark matter
models, but in cases B and C, the exponential cutoff can do so also.
This preference in some cases for the log-parabola spectrum is
predominantly coming from the lowest energy bins. The maximum
likelihood fit prefers giving no appreciable amount of photons to the
lowest energy bins, a fact that is difficult for DM models to explain,
but is more easily accommodated by the log-parabola spectra. This
preference of the lowest energy bins for the log-parabola spectra can
be seen in Figure \ref{bestfit}, where we plot the best fit models,
along with the data. It is worth noting that these lowest energy bins
have the largest systematics associated with them due to the large
size of the size of the point spread function at those energies
\cite{Calore:2014xka}.  Unlike the evidence ratios, the Bayes factors
are particularly sensitive to these systematics, particularly because
no model is a strikingly good fit, just less bad than the
others. Indeed, when ignoring the first few data points for each data
case, the Bayes factors tend to show less extreme results, giving more
consistent fits to the GCE. With these truncated data sets, the values
of the Bayes factors come from the models' abilities to explain the
difference in flux coming from the GC and the dwarfs.

On an additional note, the preference for $b\bar{b}$ can also be
seen in Fig.~\ref{bestfit}. Since the $\tau^+ \tau^-$ model
requires a light (compared to the $b\bar{b}$ model) dark matter
mass to explain the peak of the GCE spectra at around 1-2 GeV,
and since these annihilating dark matter models cutoff in energy
at around their mass, the $\tau^+ \tau^-$ model fail to account
for the GCE spectra that gradually fall off with large energies,
such as in cases A, B, and D.

\end{subsection}

\begin{subsection}{A Representative SIDM Model}

In certain classes of SIDM models for the GCE, the gamma-ray excess is
generated by the DM particles annihilating to electron-positron pairs
through a light mediator \cite{Kaplinghat:2015gha}.  The electrons
then up scatter the starlight in the Galactic Center via an IC
process.  This would naturally explain the difference in the observed
gamma-ray flux between the GC and the dwarfs since the stellar density
of the dwarfs, and therefore the interstellar light, is many orders of
magnitude smaller than the stellar mass of the GC.

Should the GCE be explained DM annihilating to electrons that interact
with the ambient starlight, the process should be governed by the
following IC equation:
\begin{equation}
\frac{d n_\gamma}{d E dt} = \sigma_T c n_e n_{\rm ISRF}
\frac{dN_\gamma}{dE}\,,
\end{equation}
where $n_\gamma$ is the number density of gamma rays, $\sigma_T$ is
the Thomson cross section, $n_e$ is the number density of electrons
produced by annihilating DM, $n_{\rm ISRF}$ is the number density of
low energy photons in the interstellar radiation field, and
$\frac{dN_\gamma}{dE}$ is the probability distribution function of
producing a gamma ray of energy $E$ via this IC process.  Naturally,
this probability distribution function depends on the probability
distribution functions of the energies of the electrons produced via
DM annihilation and the energy distribution of the starlight:
\begin{equation}
\frac{dN_\gamma}{dE} = \int dE_e dE_{\rm ISRF} p(E_\gamma|E_e , E_{\rm ISRF})
p(E_e) p(E_{\rm ISRF})\,.
\end{equation}
In principle, other energy-loss mechanisms, such as synchrotron emission,
can alter the energy distribution of electrons in this IC process.  We
checked this model against the spectrum \texttt{PPPC4DMID} calculates
and found the shape of the spectra were largely consistent.

Since the electrons are produced via a two-body interaction, the
number density of electrons should scale as the square of the number
density of DM particles: $n_e \propto n_\chi^2$.  To convert the time
derivative of the differential number density of photons to some
number flux, we need to evaluate the following integral:
\begin{equation}
\frac{d \Phi_\gamma}{dE} = \int dV' \frac{1}{4 \pi
  (\vec{R}-\vec{R}')^2} \frac{d n_\gamma}{dE dt}(\vec{R}')\,.
\end{equation}
Choosing the origin of the coordinate system to be at $R=0$ leads to
the standard expression for the J-factor, should the process be
entirely a two-body process and the time derivative of $n_\gamma$
scale solely as the square of the DM particles. Putting this
all together, we get:
\begin{equation}
\frac{d \Phi_\gamma}{dE} \propto \int d \Omega dz \frac{1}{4 \pi}
n_{\rm ISRF}n_\chi^2 \frac{dN}{dE}\,.
\end{equation}

Instead of using this equation as written, we make the following
assumptions and simplifications.  First, $n_{\rm ISRF}$ is
approximately constant where the density of DM is largest, so we can
pull the factor of $n_{\rm ISRF}$ outside the integral.  Second, it
should be true that the number density of photons from stars scales
with the stellar mass of those stars we replace $n_{\rm ISRF}$ with
the stellar mass of the gamma-ray source, $M_{*}$:
\begin{equation}
\frac{d \Phi_\gamma}{dE} \propto \frac{J}{m_\chi^2} \frac{M_*}{M_{*,{\rm GC}}}\,.
\end{equation}

Taking this spectrum leads to a model that has far greater consistency
between the GCE and dwarfs; the evidence ratios for this model are all
around unity for each of the data cases. This highlights the
possibility to alleviate the tension when going beyond simple two-body
final state scenarios.

The best fit DM mass for this representative SIDM model is $15 \pm 1$
GeV for cases A and D, $15 \pm 3$ GeV for case B, and $21 \pm 2$ GeV
for case C.

To construct a more realistic and self-consistent SIDM model, we would
need to account for two effects.  The first is Sommerfeld enhancement
in the dwarfs.  This Sommerfeld enhancement causes an increase in the
effective annihilation cross section due to the smaller velocity
dispersion in the dwarfs, relative to the GC
\cite{Kaplinghat:2015gha}.  This would tend to push down the limits on
the DM annihilation cross section coming from the dwarfs.  However,
unless this enhancement factor were many orders of magnitude above
unity, the evidence ratio would still be around unity. The second
effect would have the opposite impact on the dwarfs' cross section
limits.  Since SIDM models generically predict cored density profiles
for the dwarfs \cite{Kaplinghat:2015aga}, the inferred central density
of the dwarves would be smaller than implied by assuming an NFW
profile, as is currently done.  This, in turn, would decrease the
J-factors of the dwarfs and push up the limits on the DM annihilation
cross section.
\end{subsection}

\begin{figure*}[t!]
\begin{center}
    \includegraphics[width=3.25 truein]{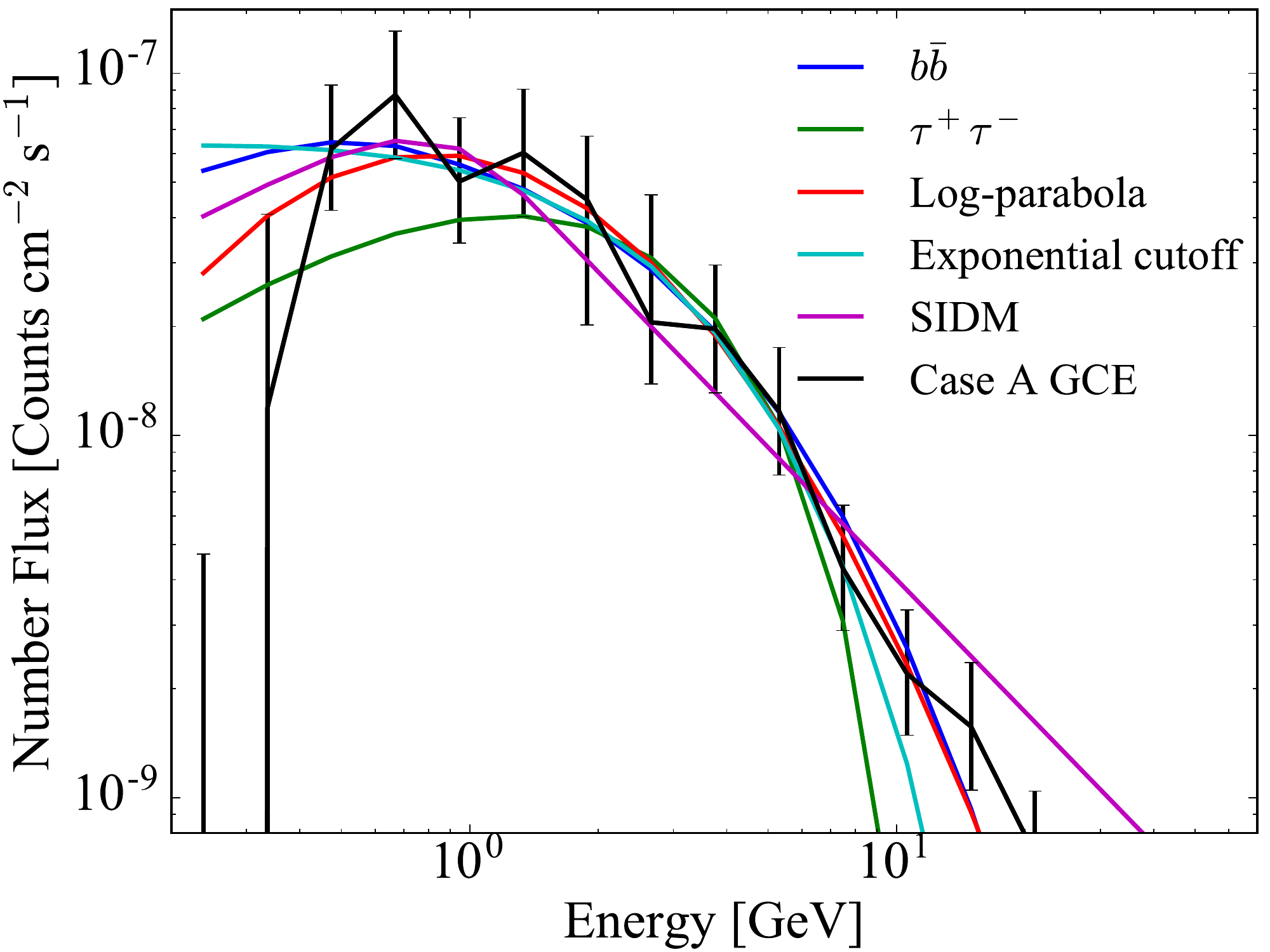}\
    \includegraphics[width=3.25 truein]{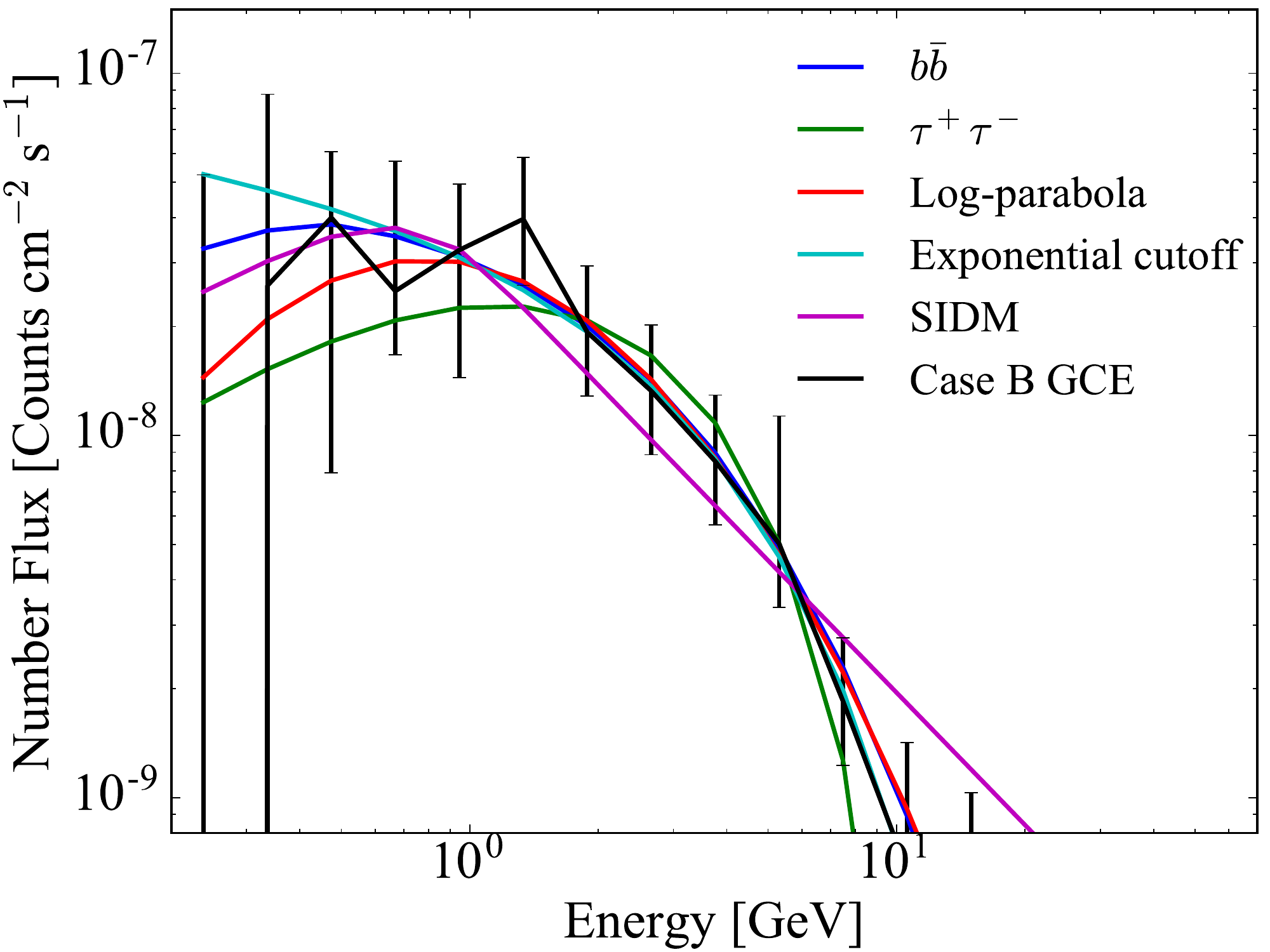}\\
    \includegraphics[width=3.25 truein]{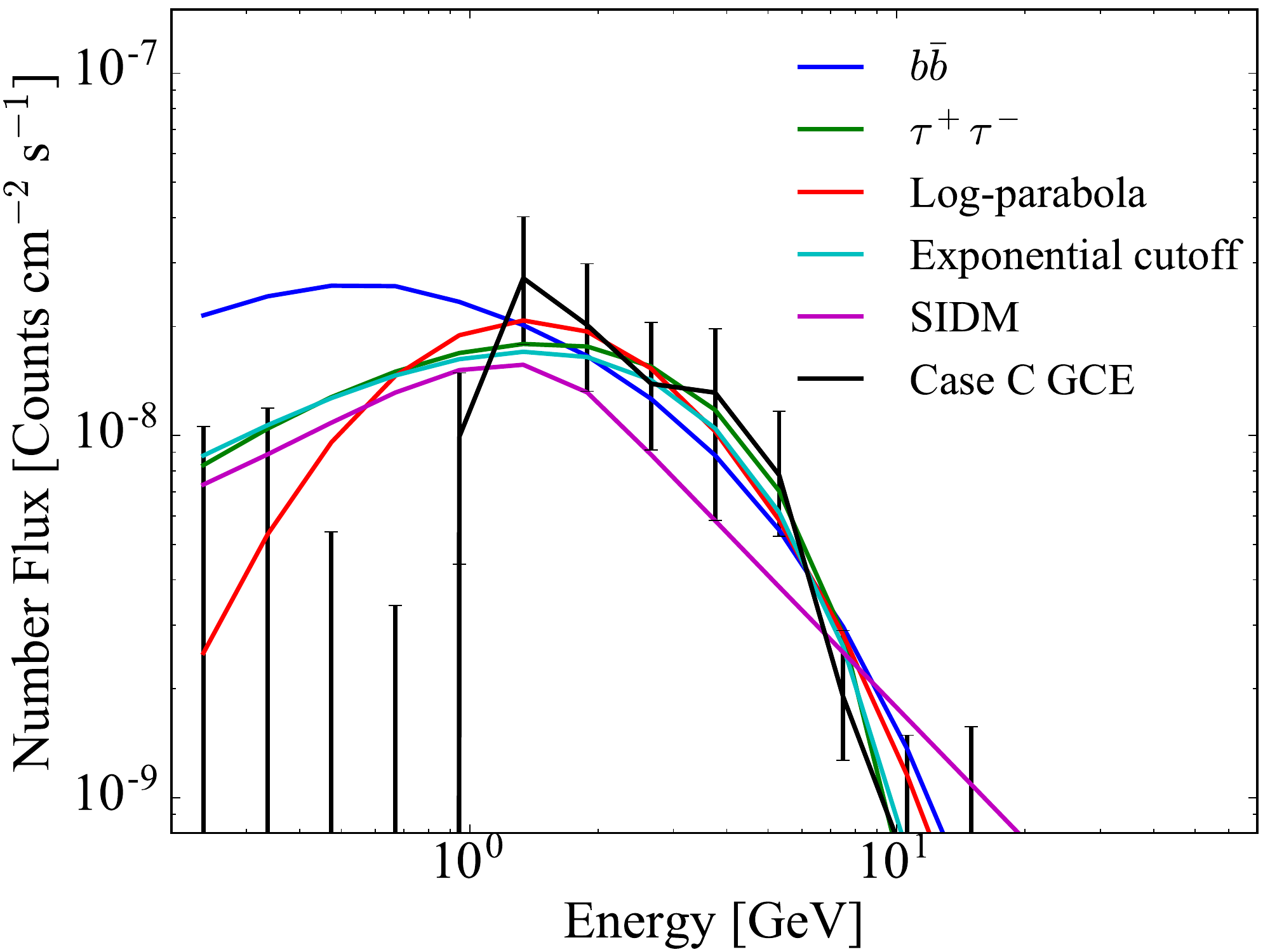}
    \includegraphics[width=3.25 truein]{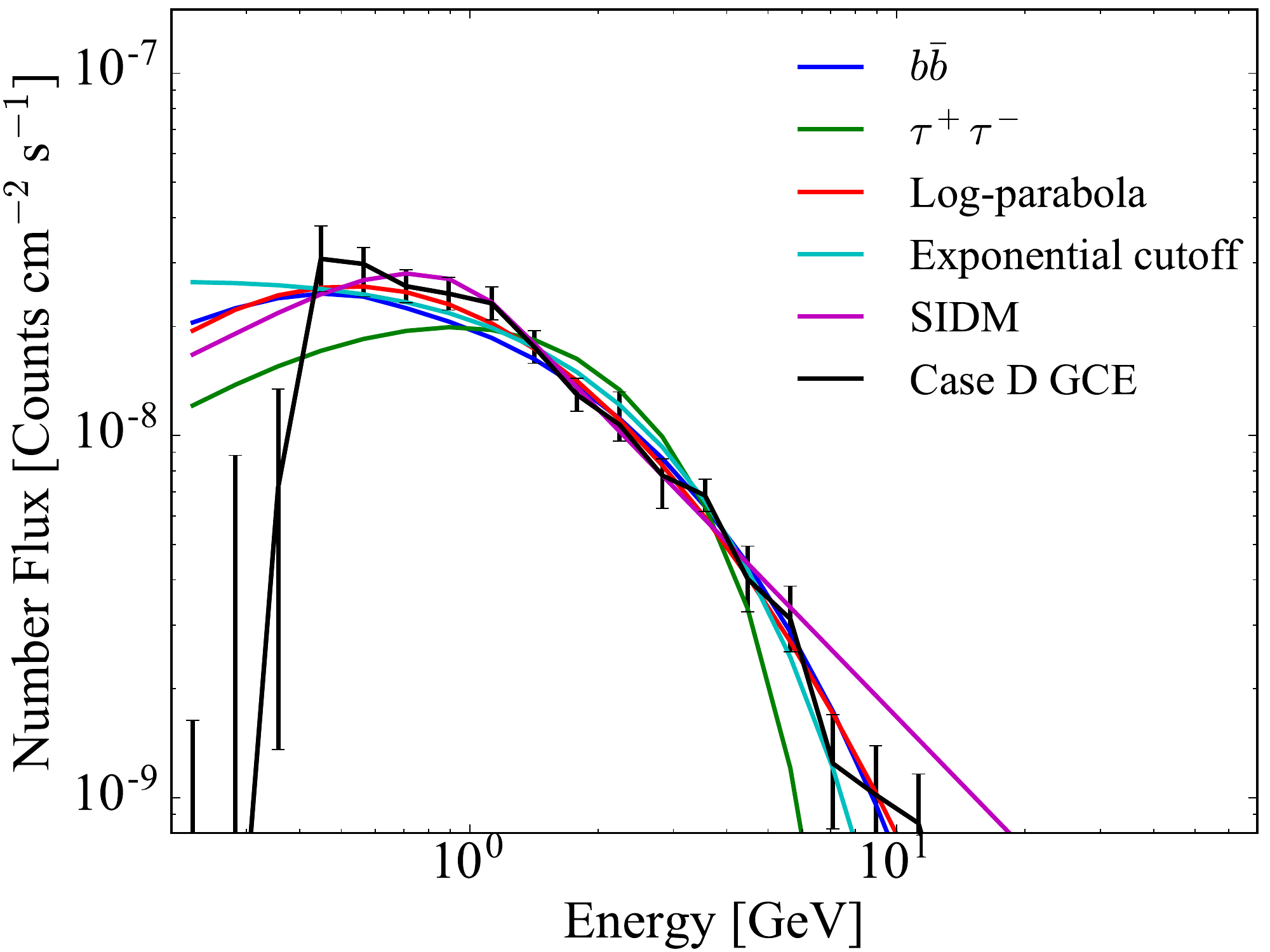}
\end{center}
\caption{Here we plot the number flux for the GCE template along
  with the best fit spectra for the different models considered.  The
  error bars correspond to the 1-$\sigma$ region of each bin's number
  flux likelihood profiles.
\label{bestfit}}
\end{figure*}

\begin{table}[]
\begin{center}
\def\arraystretch{1.5}
\caption{Bayes factors for the considered models, relative to the
  $b\bar{b}$ model, for each of the different background cases. Values
  larger than one indicate the data prefer that model over
  $b\bar{b}$.}\label{reltable}
\begin{tabular}{| l | l | l | l | l |}
 \hline
 Model                 & Case A               & Case B             & Case C              & Case D \\ \hline
 DM: $\tau^+ \tau^-$   & $4\times 10^{-24}$  & $1\times10^{-5}$    & $7\times 10^{4}$   & $1\times 10^{-22}$\\ \hline
 Log-Parabola          & $3\times 10^{12}$   & $4\times 10^{5}$    & $2\times 10^{12}$  & $5\times 10^{9}$\\ \hline
 Exponential Cutoff    & $2\times 10^{1}$    & $2\times 10^{4}$    & $4\times 10^{10}$  & 0.1\\ \hline
 SIDM                  & $5\times 10^{-20}$  & $8\times 10^{-19}$  & $6\times 10^{-2}$  & 0.1 \\ \hline
\end{tabular}
\end{center}
\end{table}

\end{section}

\begin{section}{Conclusions}
\label{conclusion}

We have analyzed the GCE in a wide variety of background models by
performing a template based likelihood analysis of the GC using four
different models for the diffuse background templates.  To answer the
question of whether an annihilating DM interpretation can be
consistent with the lack of dwarf signals, we calculated evidence
ratios for each model of the GCE and for each case of diffuse
background models.  These evidence ratios are sensitive to the choice
of background model but they all display strong to decisive tension
between the GCE and the dwarfs for annihilating DM models.
Specifically, cases A and C show decisive tension, with evidence
ratios greater than 100 for both annihilation channels, and cases B
and D show strong tension with evidence ratios greater than 10 for
both channels. This difference can, at least in part, be attributed to
the fact the likelihood fit for these cases seem to prefer both giving
less flux to the DM GCE component, and also prefer an NFW template
with a higher value for the inner slope $\gamma$.  Since the tension
is seen to various degrees using a variety of models for the the
diffuse emission, it is robust to say that prompt two-body
annihilating DM interpretations of the GCE are in strong doubt.

Astrophysical and SIDM interpretations of the GCE fare better with
evidence ratios around unity.  Ultimately, allowing the gamma ray
signal to scale with the stellar mass, as for astrophysical models, or
with the product of the J-factor and stellar mass, as with SIDM
models, relieves any tension between the GCE signal and lack of a
dwarf signal.

We also calculated Bayes factors for our different DM GCE
interpretation models.  This Bayes factor can be thought of as coming
from two different sources: the ability of the model to explain the
GCE and the ability of the model to explain the difference in GCE and
dwarf fluxes. These Bayes factors decisively prefer the log-parabola
spectrum model over the DM annihilation models in all of our
background cases, and prefer the exponential cutoff model in three of
the four background cases. This preference for either astrophysical
spectrum model predominantly comes from the lowest energy bins where
the likelihood analysis prefers to attribute no amount of flux to an
NFW template. However, these are also the energy bins that have the
largest systematic uncertainties associated with them. Standard
two-body DM annihilation models cannot explain these low energy
gamma-ray data, while more general log-parabola and exponential cutoff
models are able to do so. With the long integration time now available
from the {\em Fermi}-LAT observations of the GCE, the data allows us
to make very precise determinations of the GCE's spectral parameters,
given a particular background model. However, the accuracy of these
background models are still uncertain. In other words, the systematic
uncertainties in the background model cases dominate over the Poisson
statistical uncertainties.  In fact, there exist two sets of
tests. One is whether DM or astrophysical spectral models' can explain
the joint GCE and dwarf data. The second is the intrinsic ability of
the GCE spectral choices to explain the GC observations. Importantly,
the biggest change in the models' Bayes factors comes from the
spectral models' different ability to properly fit the GCE. In almost
all cases, log-parabola spectra is decisively better in their evidence
ratios at fitting the GCE data (cf. Table \ref{reltable}). Therefore,
given the GCE data alone, the log-parabola astrophysical
interpretation of the GCE is favored.

Furthermore, the combined GCE-dwarf data strongly to decisively
disfavor single channel DM annihilation interpretations of
the GCE. Secondary-emission from DM models like that from
SIDM could alleviate the inconsistent emission between the GCE and
dwarf galaxies. Further detailed analysis of the diffuse emission
towards the GC will help determine the true nature of the GCE and its
relation to any emission from the dwarf galaxies.

\ \\

\end{section}

\acknowledgments K.N.A. and R.E.K. are supported in part by NASA {\em
  Fermi} GI grant 15-FERMI15-0002. A.K. is supported by NSF GRFP Grant
No. DGE-1321846. N.L.R. is supported by the DOE under contracts
DESC00012567 and DESC0013999.  We thank Dan Hooper, Manoj Kaplinghat
and Savvas Koushiappas for useful discussions. We also thank Barbara
Szczerbinska and the other organizers of {\it XI International
  Conference on Interconnections between Particle Physics and
  Cosmology}, in Corpus Christi, Texas, where part of this collaborative
work was initiated.

\bibliography{master}

\end{document}